\documentclass[allclo]{FBSart}
\usepackage{epsfig,psfig}
\usepackage{amsmath}
\usepackage{wasysym}
\def\CC{\hspace*{0.15cm}\hbox{\it l\hskip -5.5pt C\/}}
\title{QED Fermi-Fields as Operator Valued Distributions and Anomalies\footnote{Presented at Light-Cone 2004, Amsterdam, 16 - 20 August}}
\author{Pierre Grang\'e $^{(1)}$  and Ernst Werner $^{(2)}$}
\institute{$(1)$ Laboratoire de Physique Math\'ematique UMR-CNRS $5825$,\\
 Universit\'e de Montpellier II,$34060$ Montpellier-Cedex, France.\\
$(2)$ Institut f\"{u}r Theoretische Physik, Universit\"{a}t Regensburg, Germany.\\ }
\runningauthor{Pierre\, Grang\'e, Ernst Werner }
\runningtitle{LC 2004}
\begin{document}
\maketitle
\begin{abstract}
 The treatment  of fields as operator valued distributions (OPVD) is recalled with the
 emphasis on the importance of causality following the work of Epstein and Glaser. 
 Gauge invariant theories demand the extension of the usual
 translation operation on OPVD, thereby leading to a generalized $QED$ formulation.
 At D=2 the solvability of the Schwinger model is totally preserved.
 At D=4 the paracompactness property of the Euclidean manifold permits using test functions which
 are decomposition of unity and thereby provides a natural justification and extension 
of the non perturbative heat kernel method (Fujikawa) for abelian anomalies. On the Minkowski
manifold the specific role of causality in the restauration of gauge invariance (and mass generation for $QED_{2}$) 
is examplified in a simple way.
\end{abstract}

\section{Introduction}
The identification of fields as operator-valued distributions (OPVD) is almost as old as Quantum
 Field Theory (QFT) itself. To cite only two early basic texts among many just recall  
Bogoliubov-Shirkov's \cite{Bogol76} and  Schweber's \cite{Schwe61} monographs. Bogoliubov and
 Shirkov were original in that their construction of the S-matrix involves test-functions 
in the range $(0,1)$ but differrent from zero only in a certain finite space-time region. The 
condition of causality applied to the supports of the test functions induces essential relations 
between the S-matrix amplitudes. This approach was later pursued and developped by Epstein and 
Glaser \cite{Epst73} leading to a well recognized causal pertubation theory which is "free of 
mathematically undefined quantities but hides the multiplicative structure of renormalization" 
\cite{Itzy80}. Apparently for most of QFT practitioners this was a fatal disease and Epstein
 and Glaser's work is scarcely referenced and almost fell into oblivion, save for the important 
contributions of Scharf \cite{Scha95}. The situation is different in the world of mathematicians 
dedicated to the construction of a rigorous and mathematically well defined QFT. In this respect
 the necessary steps involve the extension of singular distributions to the whole space-time
 manifold. An often reported conjecture made by A. Connes and independently by R. Estrada 
\cite{Estr1} states that "Hadamard's finite part theory is in principle enough to deal with QFT
 divergences". Clearly, for outsiders, the statement called for some clarifications. They came 
only recently \cite{Kuz96}. It appears that a rigorous way to get an extension of a singular 
distribution is a weighted Taylor series surgery, that is to throw away  an appropriate jet of 
the test function at the singularity. Transposed to Fourier space the procedure amounts to a 
substraction method which includes BPHZ renormalization \cite{Bogol76,BPHZ} as a special case. 
In Minkowskian metric this is equivalent to the implementation of causality while in the Euclidean
counterpart it is a symmetry preserving prescription for substractions.
The OPVD approach we implemented in the Euclidean LCQ study of the critical properties of 
$\phi^4_{1+1}$-theory \cite{SGW02} is in the line of Epstein and Glaser \cite{Epst73}. However it
 has the important additional feature that the paracompactness of the Euclidean manifold allows
 using any partition of unity as $C^{\infty}$ test functions with compact support in Fourier
 space. Thereby any UV- divergences in Fourier space integrations are properly regulated. It 
is our aim here to develop the treatment of Fermi-fields as OPVD in a gauge field environment, 
sketched at the LC$03$ Durham meeting \cite{GW03}. 
\section{Fermi-field as OPVD: problems with gauge invariant translation. Lessons from Schwinger's $QED_2$.}
Let $\psi(x)$ be the Dirac massive free field, then $(i\not{\!\!\partial} - m)\psi(x)=0$. $\psi(x)$
is an OPVD \,\, which defines a functional \, $\Phi(\rho)$ \, with respect to a test function :\\
$\Phi(\rho) \equiv <\psi,\rho> = \int d^{(D)}y\psi(y)\rho(y)$. $\Phi(\rho)$ is an operator-valued functional with the possible interpretation of a
 more general functional \, $\Phi(x,\rho)$ \, evaluated at\\ 
$x=0$. Indeed the translated functional is a well defined 
object \cite{Sch63} such that  $T_{x}\Phi(\rho) = <T_{x}\psi,\rho> = <\psi,T_{-x}\rho> = \int d^{(D)}y \psi(y) \rho(x-y)$. Due to the properties of $\rho$, $\Psi(x)\equiv T_{x}\Phi(\rho)$ obeys also Dirac's equation and 
is taken as the physical field which is now and most importantly an analytic function of the
 $x$-variable \cite{Sch63}. The possible singular behaviour of the original field  $\psi(x)$ is 
now transferred to the class of test functions necessary to define a {\it bona-fide} tempered 
extension of the original OPVD on the whole space-time manifold \cite{Epst73,Scha95,Kuz96}. For 
$QED$ the fermionic field obeys $(i\not{\!\!\partial}-e\not{\!\!\!A}- m)\psi(x)=0$, and it is
 clear that the translation in $\psi(x)$ must be done in a way compatible with gauge
transformations. The immediate candidate would be the well known Dirac string from $x$ to $y$.
 But now $\Psi(x)$ depends on the path $\gamma(s)$ from $x$ to $y$  and  does not obey exactly $QED$'s 
equation. It induces an alteration of the path integral formulation. Moreover for $QED_2$ $\Psi(x)$ does not 
exhibit the bosonization features of the known solution. However, following Dirac's early analysis \cite{Dir55}, 
 the most general writing for $\Psi(x)$ is
\begin{equation}
\Psi(x) = \int d^{(D)}y \rho(y-x) \exp [i e \int d^{(D)}z \CC^{\mu}(x,y,z) A_{\mu}(z)] \psi(y).
\end{equation}
$\Psi(x)$ transforms as the original  $\psi(x)$ under a gauge transformation provided that
\begin{equation}
\partial_{z}^{\mu}\CC_{\mu}(x,y,z)=\delta(z-x)-\delta(z-y).
\end{equation}
There $\CC^{\mu}$ may have a matrix structure. To clarify the importance of this fact consider again $QED_2$. 
In this case the basic matrices are $\mathcal{I},\gamma_{\mu},\gamma_5$. The contributions due to $(\mathcal{I} \pm
 \gamma_5)$ can be eliminated in the Lorentz gauge, for then one may look for $\CC^{\mu}$ under the form 
$\partial_{z}^{\mu} C  \rightarrow \int d^{(D)}z(\partial_{z}^{\mu}C)A_{\mu}=-\int d^{(D)}z
 C (\partial_{z}^{\mu}A_{\mu}) = 0$. Hence the relevant term involves only $\gamma_{\mu}$ and 
$\CC^{\mu}(x,y,z)=C(x,y,z) \gamma^{\mu}$. But at $D=2$ the longitudinal part of $A_{\mu}(z)$ can be gauged away,
 only the transverse part matters and one has then $\int d^{(2)}z C(x,y,z)\gamma^{\mu} A_{\mu}(z)=\int d^{(2)}z
 C(x,y,z)\gamma^{\mu}\epsilon_{\mu \nu}\partial_{z}^{\nu}\phi(z)=-\gamma_5 \int d^{(2)}z
 C(x,y,z)\gamma_{\nu}\partial_{z}^{\nu}\phi(z)=\gamma_5 \int d^{(2)}z (\partial_{z}^{\nu}
C(x,y,z)\gamma_{\nu})\phi(z)=\gamma_5 [\phi(x)-\phi(y)]$, where $Eq.(2)$ and the linearity of $C(x,y,z)$ in $\gamma$ matrices has been used. $\Psi(x)$ writes now
$\Psi(x)=\exp [i e \gamma_5 \phi(x)]\int d^{(2)}y \rho(y-x) \exp [-i e \gamma_5 \phi(y)] \psi(y)=\exp [i e 
\gamma_5 \phi(x)] \chi(x)$. This is just the bosonization ansatz of the conventional theory and it is checked
that $\Psi(x)$ still obeys $(i\not{\!\!\partial}-e\not{\!\!\!A})\Psi(x)=0$, giving the usual anomaly and mass 
generation ({\it cf} below). At $D=4$ $\CC^{\mu}(x,y,z)$ can be a $(4 \otimes  4)$ matrix built 
from $\mathcal{I},\gamma^{\mu},\gamma_5,\gamma_5\gamma^{\mu},\sigma^{\mu\nu}$. As for $D=2$ contributions from 
$\mathcal{I},\gamma_5$ can be disregarded. Clearly we don't want $\CC^{\mu}$ to mix the chiralities of the original
 $\psi(y)$ while preserving its equation of motion and gauge transformation property. This imposes a unique 
writing for  $\Psi(x)$ as 
\begin{equation}
\Psi(x)= \int d^{(4)}y\rho(y-x)[H_{+}(x,y)+H_{-}(x,y)]
\end{equation}
with
\begin{equation}
H_{\pm}(x,y)={1 \over 2} \exp [i {e \over 2} \int d^{(4)}z C(x,y,z)\gamma^{\mu}(1 \pm
\gamma_{5})A_{\mu}(z)](1 \pm \gamma_{5})\psi(y).
\end{equation}
 The solution of $Eq.(2)$ is now path independant and gives $C(x,y,z)=-i \int {d^{(4)}k \over
 (2\pi)^{4}} {\gamma.k \over k^2}[\exp(ik(z-x))-\exp(ik(z-y))]$, which cannot be gauged away 
\cite{GW03}. It is verified that  $\Psi(x)$ obeys the classical $QED$ original equation 
\footnote{The necessary commutations result from $[\gamma^{\mu}(1 \pm \gamma_5),\gamma^{\nu}(1 
\pm \gamma_5)]=0, \forall (\mu,\nu)$. Whereas for $m>0$ the distinction between right and
left-handed fermions is only technical, for $m=0$ the chirality will remain a conserved quantum
 number.} thereby permitting  a path integral formulation in terms of this "smeared" field. 
The question of charged particles and asymptotic states in gauge theories \cite{LavMc03} is now 
to be adressed with respect to this field. 
\section{Anomalies: Fujikawa's method revisited}
Using the properties of $\gamma$-matrices the right- and left-handed components can be recombined. With the variable change \, $y=x+\epsilon$ \' and \,\, $C(x,y,z)=c(z-x)\\
-c(z-y)=\epsilon.\partial_{z}c(z-x)+\mathcal{O}(\epsilon^2)$, $\Psi(x)$ reads now 
\begin{eqnarray}
\Psi(x)\!\!\!&=&\!\!\!\int {d^{(D)}p \over (2\pi)^D}\int d^{(D)}\epsilon \rho(\epsilon)
 \exp [i\epsilon.(p+e \int d^{(D)}z \partial_{z}c(z-x)\not{\!\!\!A}(z))]\exp [i p.x]\tilde{\psi}(p). \nonumber
\end{eqnarray} 
Here the neglected terms are of order $\mathcal{O}(R^2)$, where $R$ is the "small" radius of the ball, support of
 the test function $\rho(\epsilon)$. Due to rotational symmetry its Fourier transform depends on $q^2$ only and 
since $\tilde{\rho}(q^2)=\tilde{\rho}((\gamma.q)^2)$ it gives \footnote{$\tilde{\rho}(-\not{\!\!\!D}_{x}^2)$ cannot
 be pulled out of the integral since $\tilde{\psi}(p)$ is a distribution.} $\Psi(x)= \int {d^{(D)}p \over (2\pi)^D}
\tilde{\rho}((\not{\!\!p}+e \not{\!\!\!A}(x))^2)\exp [i p.x]\tilde{\psi}(p)+\mathcal{O}(R^2)=\int {d^{(D)}p \over (2\pi)^D}\tilde{\rho}(-\not{\!\!\!D}_{x}^2)\exp [i p.x]\tilde{\psi}(p)+
\mathcal{O}(R^2)$.Since the value of the mass is inessential in the sequel $m=0$ will be taken.  In the path integral formalism the measure is now expressed in terms of the {\it regularized} complete set of eigenfunctions $\{\Xi_{n} \}$ of the hermitian operator $\not{\!\!\!D}$, since $\Xi_{n}(x)$ and the unregularized $\xi_{n}(x)$ obey the same eigenvalue equation
 $\not{\!\!D}\xi_{n}(x) =\lambda \xi_{n}(x)$  by construction. Hence $\Psi(x)=\sum_{n}a_{n}\Xi_{n}(x),
\overline{\Psi}(x)=\sum_{n}b_{n}\overline{\Xi}_{n}(x)$ and $\mathcal{D}\Psi(x)\mathcal{D}\overline{\Psi}(x)=
\prod_{m,n}da_{m}d\overline{b}_{n}$. The analysis follows as usual \cite{Naka90}, save for the fact that the quantity \,\,\,
 $B(x)=\sum_{n}\Xi^{+}_{n}(x)\gamma_{5}\Xi_{n}(x)=\int {d^{(4)}p \over (2\pi)^4}tr\{\gamma_{5}\exp[-ik.x]$\\
$\tilde{\rho}^{2}(-\!\!\!\not{\!\!\!D}_{x}^2)\exp[ik.x]\}$ is now finite due to the presence of test functions. 
The paracompactness property of the Euclidean manifold implies that $\tilde{\rho}(p^2)$ can be taken as a decomposition of unity 
which introduces a scale (related to the inverse radius $R$ of the ball, support of test function in configuration space) and in fact
 $\tilde{\rho}(p^2) \rightarrow \tilde{\rho}({p^2 \over \Lambda^2})$  \cite{SGW02,GW03}. In the limit $\Lambda \rightarrow \infty$ 
$B(x)$ reduces to $B(x)={e^2 \over 8}tr[\gamma_{5}(\sigma^{\mu\nu}F_{\mu\nu})^2]\int {d^{4}y \over (2\pi)^4}
{d^2 \over d^2(y^2)}[\tilde{\rho}^{2}(y^2)]={e^2 \over 16 \pi^2} ^{\,*}F^{\mu \nu}(x) F_{\mu \nu}(x)$,
 since $\tilde{\rho}^{2}(0)=1$ (decomposition of unity). At $D=2$ only the term in $tr[\gamma_{5}\sigma^{\mu \nu}F_{\mu \nu}]$ 
survives giving also the known result. 
\section{Causality and gauge invariance in $QED_2$}
Whatever the importance of  the non-perturbative anomaly test, it is well known that any type of
 regularisation in Fourier space via damping functions  at large momenta violates pertubatively
  gauge invariance  and Ward identities. This can be avoided if causality is implemented from the
start in the construction of the S-matrix as advocated by Epstein and Glaser \cite{Epst73} and
Scharf \cite{Scha95}. This line of thought leads to use propagators with truly  causal supports
satisfaying dispersion relations being tantamount to substractions dictated by analyticity. It
guarantees finite results and gauge invariance. Moreover it is conceptually important to realize
that the well established BPHZ renormalization scheme \cite{Bogol76,BPHZ} turns out to be a
special case of the Epstein-Glaser prescription \cite{Kuz96}. For the special case of $QED_2$ 
(Schwinger model) the essential quantity is the causal polarization tensor whose form is 
\cite{Scha95,Scha98}
\begin{eqnarray}
\Pi_{\mu \nu}(k)&=&\int d^{2}x \exp [ik.x]\{Tr[\gamma_{\mu}S^{(+)}(x)\gamma_{\nu}S^{(-)}(-x)]-(\mu \leftrightarrow
 \nu, x \leftrightarrow -x)\} \nonumber\\
&=&\!e^{2}[\widehat{P}_{\mu \nu}(k)-(\mu \leftrightarrow \nu, k \leftrightarrow -k)], 
\end{eqnarray}
$S^{(\pm)}(\pm x)$  are related to the usual Feynman propagator $S^{F}(\pm x)$
corrected respectively by advanced and retarded pieces $\int {d^{2}p \over (2\pi)^2}{(\not{p}+m)\exp [-i p.x]
 \over (p^2-m^2 \mp i p^0 \epsilon)}$ to give propagators with truly causal supports. They write explicitely
\begin{equation}
S^{(\pm)}(\pm x) =\pm i \int {d^{2}p \over (2\pi)^2} 2 \pi \delta (p^{2}-m^{2})\Theta (p^{0})(\pm 
\not{\!\!p}+m)\exp [-ip.x] 
\end{equation}
With these expressions for $S^{(\pm)}(\pm x)$ and after tracing over the $\gamma$
matrices $\widehat{P}_{\mu \nu}(k)$ becomes
\begin{eqnarray}
\widehat{P}_{\mu \nu}(k)&=&-2 \int d^{2}p \delta (p^{2}-m^{2})\delta (k^{2}-2k.p)\Theta (p_0) \Theta (k_0-p_0)\nonumber\\
& &[p_{\mu} k_{\nu}+p_{\nu} k_{\mu} -2 p_{\mu} p_{\nu}-{k^2 \over 2} g_{\mu \nu}].
\end{eqnarray}
The integral over $p$ is finite and $k^{\mu}\widehat{P}_{\mu \nu}(k)=0$. It is important to note the role of the $\delta$ functions in getting this result. They are specific to the form of the propagators $S^{(\pm)}(\pm x)$. Hence $\Pi_{\mu \nu}(k)=e ^2 (g_{\mu \nu}-{k_{\mu} k_{\nu} \over k^2}) k^2 \widehat{d}(k)$ with 
\begin{equation}
\widehat{d}(k)=\lim_{m^{2} \rightarrow 0}{4 m^2 \over k^4} {1 \over \sqrt{1-{4 m^2 \over k^2}}} \Theta (k^2-4m^2)\mbox{sign}(k_0)=2 \delta(k^2)\mbox{sign}(k_0).
\end{equation}
Causality imposes redefining $k^2 \widehat{d}(k)$ through a one-time substracted dispersion relation
\begin{eqnarray}
k^2 \widehat{d}(k) \rightarrow \widehat{r}(k)&=&{i \over 2 \pi} \int^{\infty}_{-\infty}dt {\widehat{d}(tk) (tk)^2 \over (t-i\epsilon)(1-t+i\epsilon)}\nonumber \\
 &=&  {i m^2 \over \pi}\Big{[}{1\over m^2}+{2 \over k^2} {1 \over \sqrt{1-{4 m^2 \over k^2}}}\log\Big{[}{\sqrt{1-{4 m^2 \over k^2}}+1 \over \sqrt{1-{4 m^2 \over k^2}}-1 }\Big{]}\Big{]}.
\end{eqnarray}
Here $k^2 > 4 m^2, k_0 > 0$. The first term in ${1\over m^2}$ comes from the substraction itself.
Hence $\lim_{m^{2}\rightarrow 0} \widehat{r}(k)={i \over \pi}$, giving $\Pi_{\mu \nu}(k)=i{e^2 \over \pi}(g_{\mu \nu}-{k_{\mu} k_{\nu} \over k^2})$, {\it ie} the right boson mass.
In the process it is essential to identify properly the singular order of the propagator to determine the substaction
 needed in the dispersion relation since it determines completly the limit $m^2 \rightarrow 0$. If not, then UV
 divergences and violation of gauge invariance will show up.
\section{Conclusion}
The necessity of formulating Quantum Field Theory in the continuum requires  treating fields
as OPVD with  specific test functions. Thereby  a non-standard regularization scheme  is obtained which
is in the line of the Epstein and Glaser extension of singular distributions. For abelian gauge theories
  the usual translation operation on distributions has been  modified  following Dirac's early analysis
  of the possible  phase factor leading to the proper gauge transformation of the initial Fermi field itself.
The procedure  meets the necessary requirements  of leaving the solution of the Schwinger  model unaltered
 and yet still permits a path integral formulation in terms of  the smeared field. It provides naturally  
an interpretation of Fujikawa's analysis of the abelian anomaly.  Finally recognizing the filiation of
our approach with Epstein and Glaser's treatment  indicates the essential role of
causality  in restoring gauge invariance which is otherwise violated by any  regularisation with  UV damping  test
functions  only.  This  opens up the very interesting perspective of building up  a gauge invariant  LC-quantization
 framework free of divergences by construction from the outset. Only finite renormalization will occur in
connection with the intrinsic scale present  in the decomposition of unity in Fourier space or in  the
parametrization of Epstein and Glaser's weight function $w(x), w(0)=1$  in configuration space.

\begin{thebibliography}{99}
\bibitem{Bogol76} N.N. Bogoliubov, D.V. Shirkov, "Introduction to the Theory of Quantized Fields',New York, J. Wiley \& Sons, Publishers, Inc., (1957, 3rd edition 1980).
\bibitem{Schwe61} S. Schweber, "An Introduction to Relativistic Quantum Field Theory", New York, Harper \& Row, Publishers, Inc., (1961).
\bibitem{Epst73} H. Epstein, V. Glaser, Ann. Inst. Poincar\'e, {\bf A 29.} 211 (1973).
\bibitem{Itzy80} C. Itzykson, J.B. Zuber, "Quantum Field Theory", McGraw-Hill, New York (1980).
\bibitem{Scha95} G. Scharf, "Finite Quantum Electrodynamics: the causal approach", Springer Verlag, (1995).
\bibitem{Estr1} R. Estrada, R.P. Kanwal, J. Math. Anal. Appl. {\bf 141.} 195 (1989).
\bibitem{Kuz96} A.N. Kuznetsov, F.V. Tkachov, V.V. Vlasov, {\bf hep-th/9612037};
 J. Prange, J. Phys. {\bf A 32.} 2225 (1999);
 M. D\"{u}tsch, K. Fredenhagen, Com. Math. Phys. {\bf 219.} 5 (2001);
 G. Pinter, Annalen Phys. {\bf (Ser. 8) 10.} 333 (2001);
 J.M. Garcia-Bonda, Math. Phys. Anal. Geom. {\bf 6.} 59 (2003); J.M. Garcia-Bonda, S. Lazzarini, J. Math. Phys.  {\bf 44.} 3863 (2003).
\bibitem{BPHZ} W. Zimmerman, Ann. of Phys. (N.Y.) {\bf 77.} 536 (1973).
\bibitem{SGW02} S. Salmons, P. Grang\'e, E. Werner, Phys. Rev. {\bf D57.} 4981 (1998); Phys. Rev. {\bf D60.} 067701 (1999); Phys. Rev. {\bf D65.} 125015 (2002).  
\bibitem{GW03}  P. Grang\'e, E. Werner, "Light Cone Physics: Hadrons and beyond", University of Durham (UK) August 5th-9th,2003, Editor: S. Dalley (Swansea) {\bf IPPP/03/71, DCPT/03/142}.
\bibitem{Sch63} L. Schwartz, " Th\'eorie des Distributions", Hermann, Paris (1963).
\bibitem{Dir55} P.A.M. Dirac, Can. J. Phys. {\bf 33.} 650 (1955).
\bibitem{LavMc03} M. Lavelle, D. McMullan, "Light Cone Physics: Hadrons and beyond", University of Durham (UK) August 5th-9th,2003, Editor: S. Dalley (Swansea) {\bf IPPP/03/71, DCPT/03/142}.
\bibitem{Naka90} M. Nakahara,"Geometry, Topology and Physics, Graduate Student Series in Physics (1990), Institute of Physics Publishing, Bristol and Philadelphia, D.E. Brewer Editor.
\bibitem{Scha98}A. Aste, G. Scharf, U. Walther, Nuovo. Cim. {\bf A11.} 323 (1998).
\end{thebibliography}
\end{document}